\documentclass[11pt,A4paper]{article}
\usepackage{graphicx}
\usepackage{amsmath}
\usepackage{amssymb}
\usepackage{epsfig}
\usepackage[utf8]{inputenc}
\usepackage{authblk}
\usepackage{changepage}
\usepackage{color}
\usepackage{soul}

\usepackage{tabularray}
\UseTblrLibrary{amsmath}
\usepackage{relsize}
\usepackage{scrextend}
\newcommand\BibTeX{{\rmfamily B\kern-.05em \textsc{i\kern-.025em b}\kern-.08em
T\kern-.1667em\lower.7ex\hbox{E}\kern-.125emX}}
\usepackage[hidelinks]{hyperref}
\usepackage{setspace}
\usepackage{booktabs}
\usepackage[authoryear]{natbib}
\usepackage{multirow}
\usepackage{array}
\usepackage[margin=2cm]{geometry}

\newcommand{\aifsdd}{Anemoi-D$^2$}

\title{Downscaling weather forecasts from Low- to High-Resolution with Diffusion Models}
\author{Joffrey Dumont Le Brazidec, Simon Lang, Martin Leutbecher, Baudouin Raoult, Gert Mertes, Florian Pinault,  Aristofanis Tsiringakis, Pedro Maciel, Ana Prieto Nemesio, Jan Polster, Cathal O’Brien, Matthew Chantry
}

\affil{European Centre for Medium-Range Weather Forecasts (ECMWF)}
\date{December 2025}

\begin{document}

\maketitle

\begin{abstract}

We introduce a probabilistic diffusion-based method for global atmospheric downscaling implemented within the Anemoi framework.
The approach transforms low-resolution ensemble forecasts into high-resolution ensembles by learning the conditional distribution of finer-scale residuals, defined as the difference between the high-resolution fields and the interpolated low-resolution inputs. 
The system is trained on reforecast pairs from ECMWF’s IFS, using coarse fields at 100 km to reconstruct fine-scale variability at 30 km resolution. 
The bulk of the training focuses on recovering small-scale structures, while fine-tuning in high-noise regimes enables the generation of extremes.
Evaluation against the medium-range IFS ensemble target shows that the model increases probabilistic skill (FCRPS) for surface variables, reproduces target power spectra at small scales, captures physically consistent multivariate relationships such as wind–pressure coupling, and generates extreme values consistent with those of the target ensemble in tropical cyclones.

\begin{figure}
    \centering
    \includegraphics[width=0.5\linewidth]{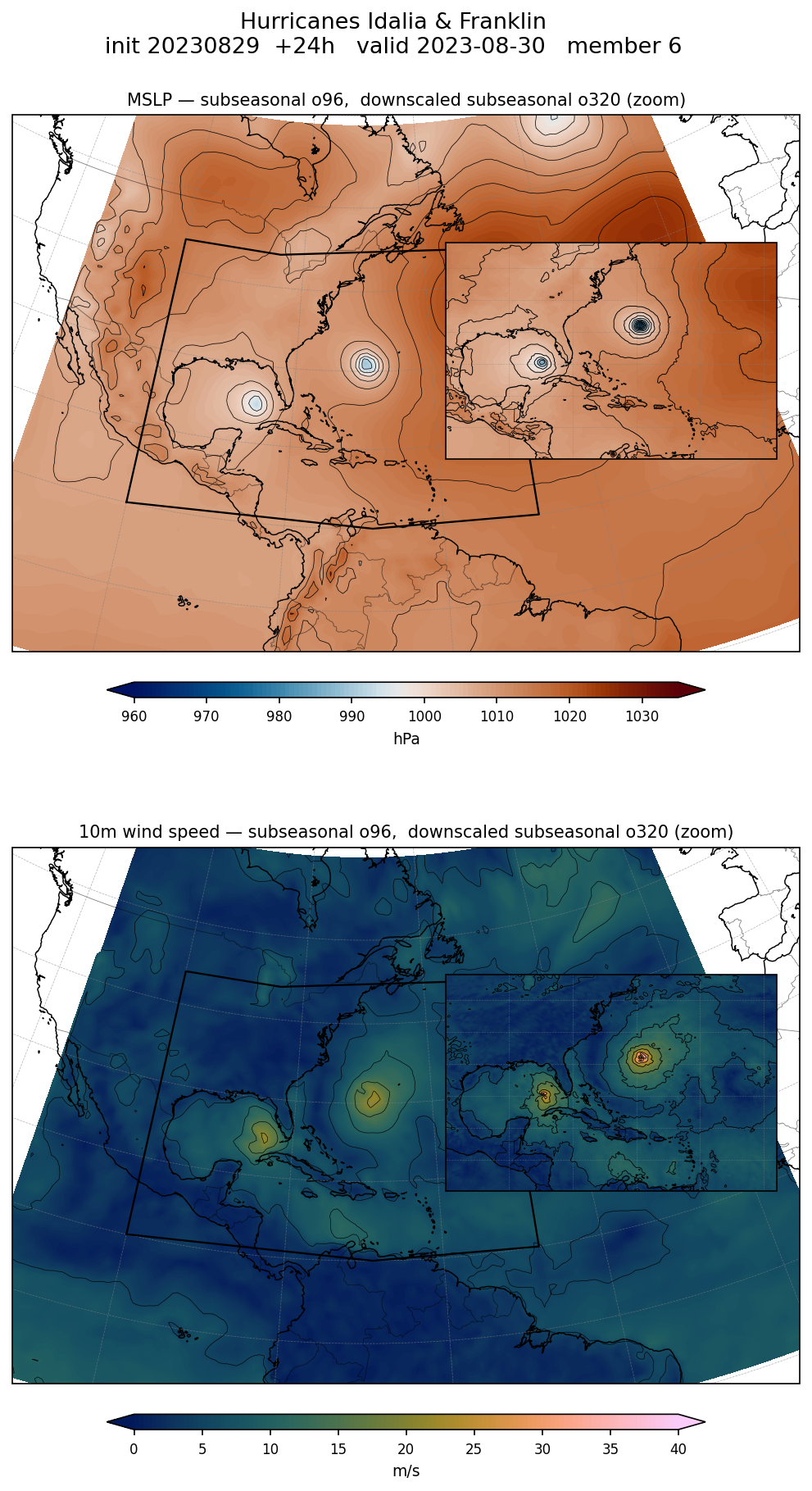}
    \caption{Hurricanes Idalia and Franklin as seen in the subseasonal ensemble forecast (eefo, O96 resolution) and its
  AI-based downscaling to O320 resolution. The background field in each panel shows the low-resolution O96 input,
   while the inset (black frame) shows the corresponding downscaled O320 prediction from the diffusion model.
  Top: mean sea level pressure (hPa). Bottom: 10-metre wind speed (m/s). Initialisation date: 29 August 2023, +24
   h lead time (valid 30 August 2023), ensemble member 6. The downscaled fields resolve the tight inner-core
  structure of both tropical cyclones — including the pressure minima and eyewall wind maxima — that is absent
  from the coarse input.}
    \label{fig:abstract}
\end{figure}

\end{abstract}

\section{Introduction}
\label{intro}

Over the last few years, several data-driven weather prediction models have achieved performance comparable or superior to that of traditional physics-based models on several standard forecast scores at coarse resolutions ($\sim$ 100–30 km) \citep{keisler_forecasting_2022, bi_pangu-weather_2022, lam_graphcast_2023, price_gencast_2024, lang_aifs_2024, lang_aifs-crps_2024}. 
However, a native medium-range km-scale high-resolution global forecasting system has yet to be achieved. 
High spatial resolution is essential to represent extremes such as intense precipitation, or tropical-cyclone peak winds.
When small-scale features influence predictive skill, direct high-resolution forecasting is justified. 
In many cases, though, large-scale dynamics dominate, and high-resolution forecasts are mainly needed locally, for diagnostic purposes. 
In such cases, it is computationally more economical to run forecasts at low resolution and apply downscaling only where and when high-resolution information is required.
Indeed, because downscaling relies primarily on local spatial information, it can reconstruct fine-scale variability without resolving the full global dynamics, making it especially suitable for computationally demanding applications such as ensemble forecasting and climate simulations.
This logic underpins approaches such as cBottle, which performs low-resolution forecasting followed by on-demand high-resolution downscaling of climate simulations \citep{brenowitz_climate_2025, Perkins2025HiROACEFA}.

Several approaches have been developed for downscaling.
Dynamical downscaling couples kilometre-scale regional models to coarser global models, allowing explicit simulation of local processes but at very high computational cost.
Traditional statistical downscaling, by contrast, is computationally efficient but can distort or even suppress meaningful climate signals, leading to biased or misleading projections \citep{chandel_state---art_2024}.

Downscaling aims to recover small-scale features from coarse-resolution inputs. 
Here, small scales refer to the scales that are not fully resolved by the coarse resolution input. 
What happens on the small scales  is not completely determined by the input. There is no correct single response to a given coarse resolution input. It is more appropriate to formulate the problem as that of finding a probability distribution of full resolution fields conditioned on the coarse resolution input.
Deterministic models trained to minimise the mean squared error approximate the mean of this conditional distribution, resulting in smoothed and less realistic fields. Here, we are interested in skilful full resolution ensemble forecasts.
To correctly represent the full variability, it is therefore necessary to construct an ensemble of downscaled fields, where large-scale uncertainty estimates originate from the low-resolution ensemble and small-scale uncertainty estimates are generated by the probabilistic downscaling framework.
The overall aim of this study is to explore machine-learned  downscaling for the generation of high-resolution ensemble forecast from inputs consisting of low-resolution ensemble forecast.

This requires a generative probabilistic framework capable of approximating the probability distribution of the high-resolution data. 
Several machine learning approaches have been explored for this purpose. 
Generative Adversarial Networks (GANs) have shown promise, for example for precipitation forecasting \citep{harris_generative_2022, harder_rainshift_2025}. 
CRPS-based methods offer an alternative for probabilistic downscaling \citep{lang_aifs-crps_2024, schillinger_enscale_2025}. 
Diffusion models have emerged as powerful and stable probabilistic generators for downscaling applications \citep{addison_downscaling_2022, mardani_residual_2023, wan_regional_2025}.

In this work, we adapt the AIFS diffusion implementation within the Anemoi framework \citep{lang_aifs_2024, nipen2024regionaldatadrivenweathermodeling,wijnands2025comparisonstretchedgridlimitedareamodelling} for downscaling applications and provide the methodological framework  \aifsdd  (DownscalingDiffusion) 
The method is applied to generate 30-km ensemble forecasts from 100-km ensemble inputs. This provides a controlled test bed to develop and validate the downscaling methodology before extending it to kilometre-scale global ensembles.

A central challenge in statistical downscaling lies in pairing input and target fields that are spatially consistent. 
Accurate alignment at the coarse scale is essential to ensure that the model learns the small-scale residual structures rather than large-scale uncertainties and biases. 
Alternative approaches, such as \cite{wan_regional_2025}, address this issue using flow matching to transform one distribution (e.g. ERA5) into another, while others train regional models forced by global models (GCM–RCM frameworks).
Here, we train on ECMWF IFS reforecasts, with lower-resolution subseasonal forecasts as inputs and higher-resolution medium-range forecasts as targets, both originating from the same initial conditions, ensuring reasonable consistency between pairs of forecasts at the two resolutions.
The method is directly applicable to both regional and global downscaling tasks.

Section~\ref{dataset} introduces the dataset and experimental setup. 
Section~\ref{model:nn} outlines the graph neural network architecture derived from AIFS \citep{lang_aifs_2024}, while Sections~\ref{model:residuals},~\ref{model:diffusion}  and~\ref{model:noise} describe the diffusion formulation and the residual-prediction strategy. 
The training and sampling procedures are presented in Section~\ref{model:training}.
Model performance is then evaluated by generating 30-km ensemble forecasts from 100-km ensemble inputs. 
We assess probabilistic skill  (Section~\ref{res:skill}), physical realism through spectral analysis and local regional diagnostics (Section~\ref{res:spectra}), and the representation of extremes through tropical-cyclone case studies (Section~\ref{res:TC}).

\section{Dataset}
\label{dataset}

The downscaling model is trained on retrospective forecasts (hindcasts), that is, forecasts initialised from past observed conditions and used to assess model skill for specific past weather situations. As input and target we use ensemble control forecasts, \textit{i.e.}\  the unperturbed reference forecast within the IFS ensemble \citep{molteni1996ecmwf}. The input consists of the  IFS sub-seasonal control hindcast with a  native resolution grid O320\footnote{Naming convention for the octahedral reduced Gaussian grid: O$N$ with $N$ the number of latitudes between the equator and the pole.} (about 30 km),  coarsened to O96 (about 100 km). The corresponding target is the IFS medium-range  control hindcast with a native resolution of O1280 coarsened to O320 (about 30 km). This coarsened setup provides a computationally efficient framework to validate the downscaling strategy before scaling up to the native resolutions. At each training step, the model receives a low-resolution representation of the atmospheric state at time  $t$ and is trained to reconstruct the corresponding high-resolution state at the same time  $t$. Training is performed exclusively on matched pairs of low- and high-resolution control forecasts; perturbed ensemble forecasts are only used at inference time, when generating high-resolution ensembles.

The dataset is post-processed with a Finite Element-based regridding over structured triangular meshes supported by linear shape functions, therefore a linear interpolation. Other interpolation methods are available to study (\emph{eg.} mesh-based bilinear, stencil-based bilinear, cubic and quasicubic, and area-conservative) however currently not considered. The regridding operation is neither conservative nor are masking, limiters or filters applied. The following regridding operators are in use: upscaling (coarsening) $U_{1280\to320}$, $U_{320\to96}$ and downscaling (refining) $D_{96\to320}$, based on linear interpolation.

Hindcast inputs and targets are initialised from the same initial analysis, which ensures exact pairing at the initial time of the forecast. As the lead time increases, the input and output hindcasts gradually diverge due to model differences. This raises the question of the optimal forecast range to use for training.
If too few lead times are included, the dataset is too small for effective training. 
Conversely, at large lead times, the input and target forecasts diverge and the model risks focusing on correcting large-scale differences (representing uncertainties at larger scale and biases) rather than capturing the fine-scale details. 
Choosing an intermediate forecast range ensures sufficient training data while preserving meaningful input-output pairs to focus the training on  capturing  the desired small-scale features.
Following a Goldilocks principle, this balance is achieved by using lead times from T+0 to T+72 hours, sampled every 12 hours. The 12-hour interval is imposed by the temporal resolution available in the archived hindcasts, rather than by the training methodology, and finer sampling could be used if such data were available.

The resulting dataset contains a total of 20,160 forecast samples. 
Model reference dates fall on Mondays and Thursdays between 29 June 2023 and 11 November 2024, yielding 144 dates in total. For each model reference date, forecasts are available at 7 lead times, from 0 to 72 hours in 12-hour increments, giving 1,008 samples per year. This set of dates and lead times is repeated for 20 consecutive years between 2003 and 2023, resulting in a total of 20,160 samples across the full period.
The O320 high-resolution dataset totals 1.35 TiB for all 20,160 forecast samples.
The training period covers 2003–2022, and 2023 is reserved for validation.

The input and output fields of  \aifsdd{}  include a range of atmospheric variables at the surface and on pressure levels, and are similar to those of AIFS, as shown in Table  \ref{dataset:fields:tab}.

\begin{table*}\centering
\begin{tabular}{>{\raggedright\arraybackslash}p{4.5cm} 
                >{\raggedright\arraybackslash}p{3.5cm} 
                >{\raggedright\arraybackslash}p{3.5cm} 
                >{\raggedright\arraybackslash}p{4cm}}
\toprule
\textbf{Field / Weather variable} & \textbf{Level} & \textbf{Input/Target} & \textbf{Normalisation} \\
\midrule
Geopotential, horizontal and vertical wind components, temperature, specific humidity 
& Pressure levels: 50, 100, 200, 300, 400, 500, 700, 850, 925, 1000\,hPa 
& Input: low-res \newline Target: high-res 
& Standardised, except geopotential (max-scaled) \\
\midrule
10m wind (u, v), 2m temperature, 2m dewpoint temperature, surface pressure, skin temperature, mean sea level pressure, total column water 
& Surface 
& Input: low-res \newline Target: high-res 
& Standardised \\
\midrule
Land–sea mask, orography, cosine/sine of local time of day, day of year, latitude, longitude, insolation, 
& Surface 
& Input: high-res 
& Not normalised \\
\bottomrule
\end{tabular}
\vspace{1mm}
\caption{Meteorological fields, levels, and normalisation used for training \aifsdd{}.}
\label{dataset:fields:tab}
\end{table*}

For model evaluation, we use the real-time sub-seasonal ensemble forecasts in 2024, coarsened to O96, as input. The diffusion model is applied independently to each member to generate the corresponding high-resolution ensemble forecast.

\section{Model}
\label{model}

\subsection{Neural network}
\label{model:nn}

\begin{figure}
    \centering
    \includegraphics[width=1\textwidth]{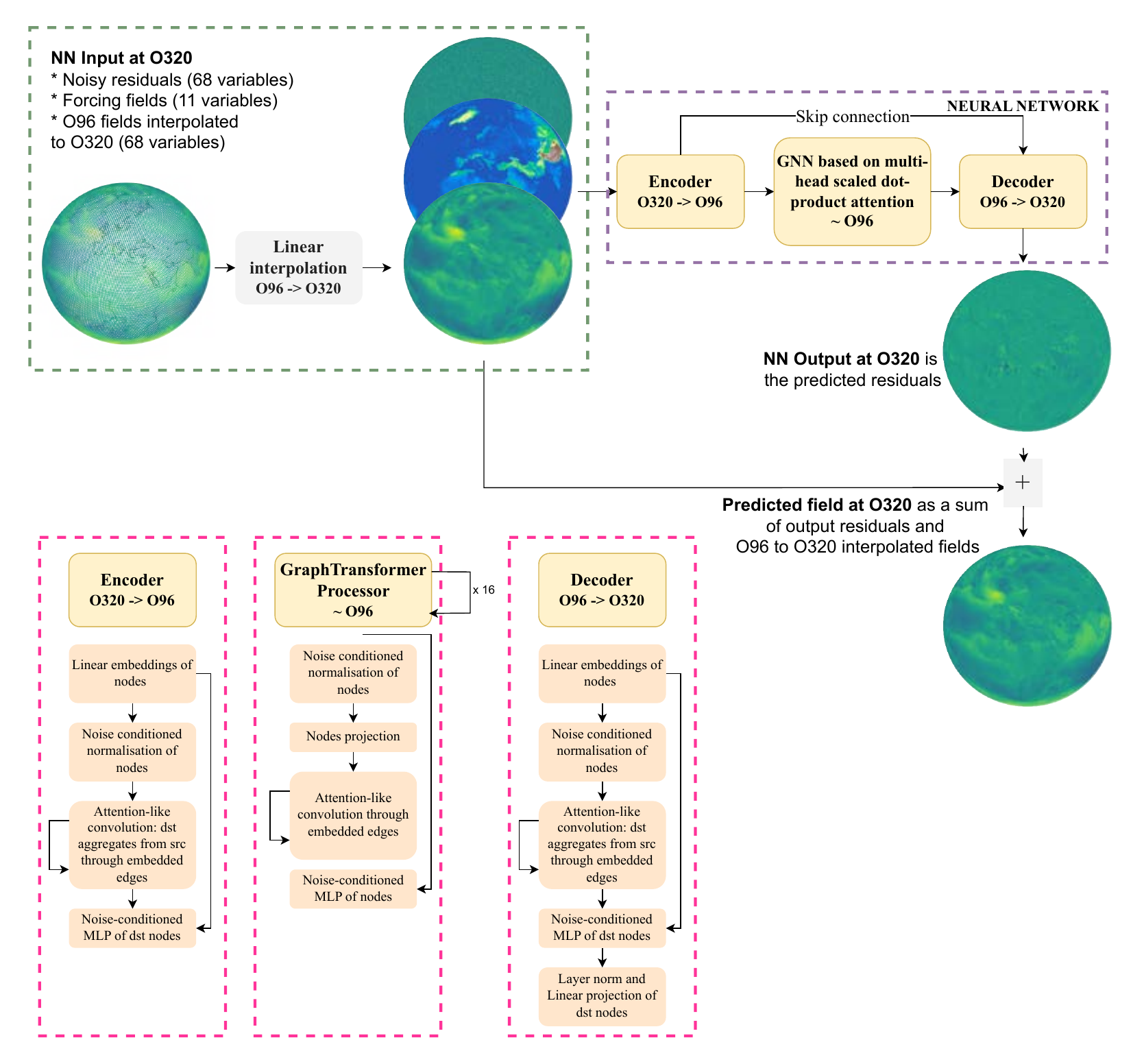}
    \caption{Schematic of the diffusion-based downscaling model. At each denoising step, the model takes as input on the O320 grid: (i) the current noisy residual estimate, (ii) static and temporal forcing fields, and (iii) the coarse O96 atmospheric fields interpolated to O320 ($D_{96\to320}$). These features are processed by an encoder–processor–decoder graph neural network: the encoder projects information from O320 to the hidden grid, the processor applies 16 successive graph-transformer layers on the hidden graph, and the decoder projects the result back to O320. In the lower panel, the pink boxes represent these three consecutive components. The model output is a prediction of the residual on O320, which is added to the interpolated coarse field to reconstruct the final high-resolution field.}
\label{model:nn:fig}
\end{figure}

The \aifsdd{} model follows a graph encoder–processor–decoder architecture (AIFS), shown in Fig. \ref{model:nn:fig}.
Its encoder and decoder are transformer-based graph neural networks (GNNs).

\begin{figure}
    \centering
    \includegraphics[width=1\textwidth]{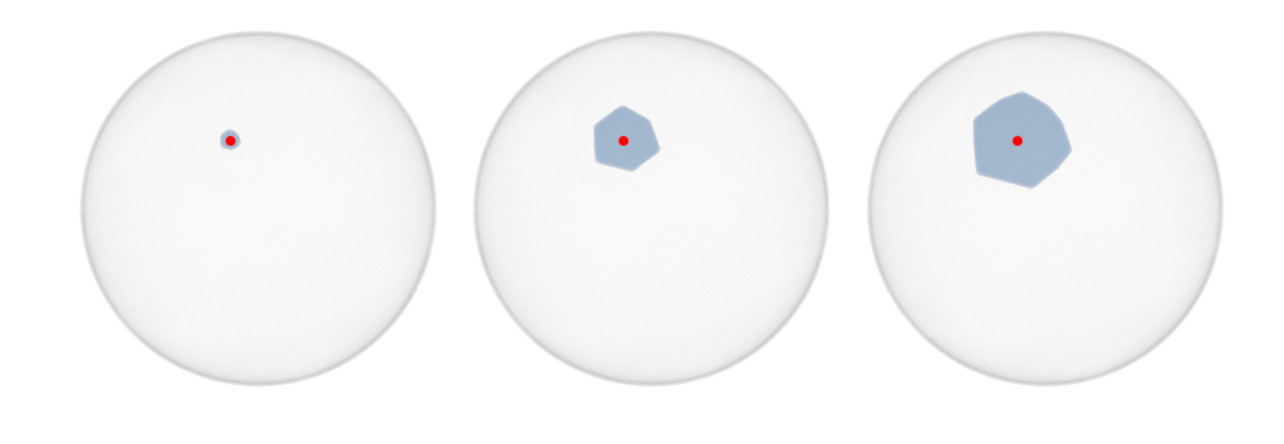}
    \caption{
Receptive field of the GNN processor for a randomly selected node (red). The figure shows how many nodes receive its information after 2, 10, and 16 message-passing layers—that is, across the corresponding 2-, 10-, and 16-hop neighborhoods. The graph structure is fixed; only the range of information propagation increases.    
    }
\label{model:nn:proc:fig}
\end{figure}

The processor is also a transformer-based graph neural network. The graph nodes are arranged on a triangularly refined icosahedron, yielding 40,962 nodes—comparable to the O96 reduced Gaussian octahedral grid. 
To keep the notation simple, Fig. \ref{model:nn:fig} describes the processor grid as $\sim$ O96.
Edges connect processor nodes with a 1-hop distance, consistent with the local nature of downscaling, where each point primarily depends on nearby spatial information.
The resulting receptive field of processor nodes after 2, 10, and 16 layers is shown in Fig. \ref{model:nn:proc:fig}.

All three components, encoder, processor, and decoder, are based on noise conditioned normalisation of the nodes, linear embeddings and attention-like convolution through embedded edges type layers as shown more precisely in Fig. \ref{model:nn:fig}.
The processor has 16 layers, an embedding dimension of 512, and 8 attention heads, resulting in a total of  62.3 million parameters. 
Finally, the model uses 147 input channels: 68 for the interpolated inputs, 68 for the noisy target during training (or the current noisy sample during inference), and 11 for high-resolution forcings.

\subsection{Residuals prediction}
\label{model:residuals}

The diffusion model is trained to predict the residuals between the target weather variables and the input interpolated onto the high-resolution grid. 
We use residuals computed through simple interpolation rather than from a fully deterministic model. 
Approaches such as \cite{mardani_residual_2023}, which rely on a deterministic model to generate residuals, risk overfitting: the residuals encountered during inference can differ from those in training.


\subsection{Diffusion model}
\label{model:diffusion}

The model follows the diffusion paradigm and, hence, the training is probabilistic.
A diffusion model defines a forward Stochastic Differential Equation (SDE) that gradually adds noise to data.
The objective is to learn the reverse SDE, which maps noise samples back to the data distribution, here corresponding to the high-resolution forecast residuals. 
This is done by parameterising the reverse dynamics with a neural network trained on the data distribution. 
A very good introduction on diffusion models is given by \citep{luo_understanding_2022}.
The \aifsdd{} diffusion model is based on AIFS-Diffusion \citep{aifs_ensemble_blog}, and follows the principles of EDM \citep{karras_elucidating_2022}.

Our goal is to generate residuals $r=y-\tilde{x}$, where $y$ is the high-resolution target and $\tilde{x}$ the interpolated coarse input field, by sampling from the conditional distribution $p(r \mid x)$.
To do so, the denoiser $D_\theta(r + n; \sigma; x)$ is trained to predict the expected residual r using the objective
\begin{equation}
    \min_{\theta} \; \mathbb{E}_{(x, r) \sim p_{\text{data}}} \; \mathbb{E}_{\sigma \sim p_{\sigma}} \; \mathbb{E}_{n \sim \mathcal{N}(0, \sigma^2 I)}  \, \| D_{\theta}(r + n; \sigma; x) - r \|^2 \, 
\label{loss_diffusion}
\end{equation}

where the noise $n$ samples a multivariate normal distribution $\mathcal{N}(0, \sigma^2 I)$ and $\theta$ denotes the neural network parameters. The denoiser is conditioned on the coarse-resolution input, high-resolution forcing fields, and the noise level $\sigma$.
It follows EDM design principles: it is explicitly conditioned on the noise level 
$\sigma$ and preconditioned to ensure stable behaviour across noise scales by controlling input and output magnitudes. Furthermore, depending on the noise level, the neural network predicts either the signal, the noise, or an intermediate representation.

At inference time, samples from $p(r \mid x)$ are generated using the EDM sampler, which integrates the reverse SDE with a predictor-corrector (Heun) scheme and introduces additional stochasticity through churning to increase sample diversity. 
Each sampled residual $r$ is then added to the interpolated coarse field to high resolution $\tilde{x}$, yielding the final high-resolution prediction $\hat{y} = \tilde{x} + r$.

\subsection{Noise levels in the diffusion model}
\label{model:noise}

Diffusion can be interpreted as spectral autoregression \citep{dieleman_diffusion_2024}: the noise level determines which spatial scales are being reconstructed.
At high noise levels, the diffusion process primarily recovers large-scale structures.
At low noise levels, the large-scale components are already visible: the denoising model reconstructs fine-scale features.
In our downscaling framework, the subseasonal input already provides the large-scale structure.
Our objective is therefore to design a model to refine this input by adding the missing small-scale variability rather than correcting large-scale biases.

For this purpose, one option is the lognormal noise distribution of EDM \citep{karras_elucidating_2022} with mean $-1.2$ and standard deviation $1.2$.
Adapted for images, it generates on weather fields (characterised by many spatial points and variables) relatively low average noise.
This primarily obscures small-scale features and drives the model to focus its learning on their reconstruction.

At 30 km resolution, the global domain contains 421\,120 grid points and 68 variables. 
Cross-grid and cross-variable correlations increase with resolution and dimensionality, so useful signal persists under substantial noise.
In our case, under very high noise variance, the MSE on normalised residuals remains lower than that obtained when training on pure noise of equal variance, confirming that meaningful structure persists despite heavy noise.

This has implications for both inference and training.
Typically, certain patterns—whether extreme events or other coherent structures—remain statistically identifiable under substantial noise. If, during training, the model is not exposed to high enough noise levels to fully destroy these structures, it never learns how such patterns re-emerge from a pure noise state.

This matters at inference as well. If sampling starts from a noise level that is too low, the earliest denoising steps take place in a regime where the model expects some physical structure to already be present. However, when initialisation is from pure noise, any apparent structure at that stage is only a random fluctuation. The model may then interpret this random Gaussian structure as meaningful signal and refine it, rather than generating the correct large-scale patterns from scratch.
Initialising (and training) at very high noise instead ensures that all structures are fully erased at the start. Coherent features—including extremes—must then emerge progressively along the denoising path. 
Thus, we have to design a training strategy that enables the downscaling model to focus on reconstructing small-scale features, corresponding to low-noise conditions, while remaining sufficiently exposed to high-noise regimes to ensure robust generation. We therefore use the two-stage training procedure summarized in Table~\ref{tab:training_stages}.

At inference time, sampling is performed with $40$ denoising steps following the \cite{karras_elucidating_2022} sampling scheduler, similar to \cite{price_gencast_2024}, but with substantially higher maximum noise levels.
The trajectory starts from a maximum noise level of $\sigma_{\max} = 10000$ and is progressively reduced to $\sigma_{\min} = 0$, with $\rho = 7$.

\begin{table}[t]
\centering
\caption{Summary of the two-stage training procedure.}
\begin{tabular}{p{1.5cm}p{2cm}p{5.5cm}p{4.5cm}}
\hline
Stage & Steps & Noise setup & Objective \\
\hline
1 & $10^6$ &
EDM lognormal noise distribution \citep{karras_elucidating_2022} with mean $-1.2$ and standard deviation $1.2$
& Base training \\

2 & $10^5$ &
Higher-noise fine-tuning with a Karras-style schedule, using $\rho = 7$, $\sigma_{\max} = 10000$, and $S_{\max} = 10000$
& Improve generation from strongly degraded states and better capture extremes \\
\hline
\end{tabular}
\label{tab:training_stages}
\end{table}

\subsection{Training}
\label{model:training}

The loss function of the diffusion model is an area-weighted mean squared error (MSE) between the predicted and target atmospheric residual fields, with variable-specific scaling. The scaling was chosen empirically to prioritise surface variables with abundant small-scale features over pressure-level variables. Similarly to \cite{lang_aifs_2024, lang_aifs-crps_2024}, loss weights also decrease linearly with height, so upper-atmosphere levels contribute less to the total loss.
AdamW \citep{loshchilov_decoupled_2019} is used as the optimizer with $\beta$-coefficients set to 0.9 and 0.95, and a weight decay setting of 0.1.

The first training phase comprises a total of 1,000,000 iterations (parameter updates), with an initial learning rate of $10^{-3}$.
We use a cosine schedule with 1000 warm-up steps, during which the learning rate increases linearly from zero to the initial learning rate. The learning rate is then reduced from its maximum value to zero.
The second phase consists of 100,000 iterations with an initial learning rate of $10^{-6}$ and a similar schedule.

The training process is distributed across 4 NVIDIA A100 64\,GB GPUs, using data parallelism. We use a total batch size of 4. The total training time corresponds to approximately 1920 GPU-hours.


\section{Results}
\label{res}

We generate a \(30\,\mathrm{km}\) ensemble forecast from a 10-member \(100\,\mathrm{km}\) ensemble forecast by applying the diffusion model independently to each ensemble member every 24h up to 10 days.

\subsection{Skill}
\label{res:skill}

\begin{figure}
    \centering
    \includegraphics[width=1\linewidth]{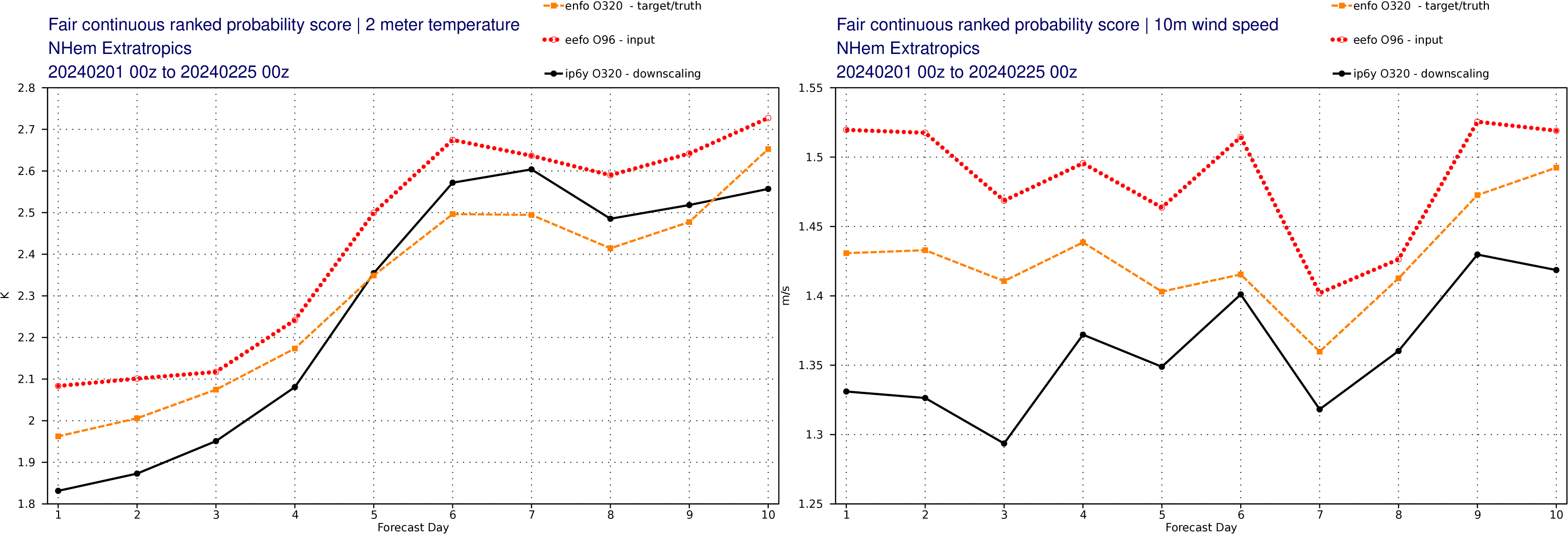}
    \caption{Fair Continuous Ranked Probability Score (FCRPS) for 2-metre temperature (left) and 10-metre wind speed (right) over the Northern Hemisphere extratropics, from forecast day 1 to 10 (20240201–20240225), evaluated against SYNOP station observations. Comparison between the downscaled ensemble forecast, the IFS medium-range ensemble forecast, and the subseasonal ensemble forecast (input).}
    \label{res:skill:fcrps:fig}
\end{figure}

Figure~\ref{res:skill:fcrps:fig} compares the probabilistic forecast skill of the input subseasonal ensemble forecast, the target or reference IFS medium-range ensemble forecast, and the ensemble obtained by downscaling the subseasonal input. For both temperature and wind speed, the downscaled forecasts exhibit systematically lower FCRPS values than the input, indicating improved probabilistic skill.

For surface variables, two factors may explain why the FCRPS of the downscaled ensemble remains different from that of the target physics-based ensemble. First, the generated ensemble has a larger spread than the target ensemble. Second, the diffusion-based downscaling model and the IFS medium-range ensemble do not represent uncertainty in the same way. Because it is trained on paired control forecasts, the downscaling model learns the conditional probabilistic structure of the target from variability present in the training dataset. In contrast, the spread of the IFS medium-range ensemble arises from explicit perturbation methods, including perturbations to the initial conditions and a stochastic representation of model uncertainty based on singular vectors and an ensemble of data assimilations \citep{leutbecher2017, lang2021}. These perturbation mechanisms are not represented in the training setup of the downscaling model. 

For pressure-level fields, by contrast, FCRPS shows no improvement relative to the subseasonal input, indicating that the model does not meaningfully correct the large-scale structure. Indeed, this is not the objective of the downscaling approach presented here.

 
\subsection{Physical realism and  power spectra}
\label{res:spectra}

\begin{figure}
    \centering
    \includegraphics[width=1\linewidth]{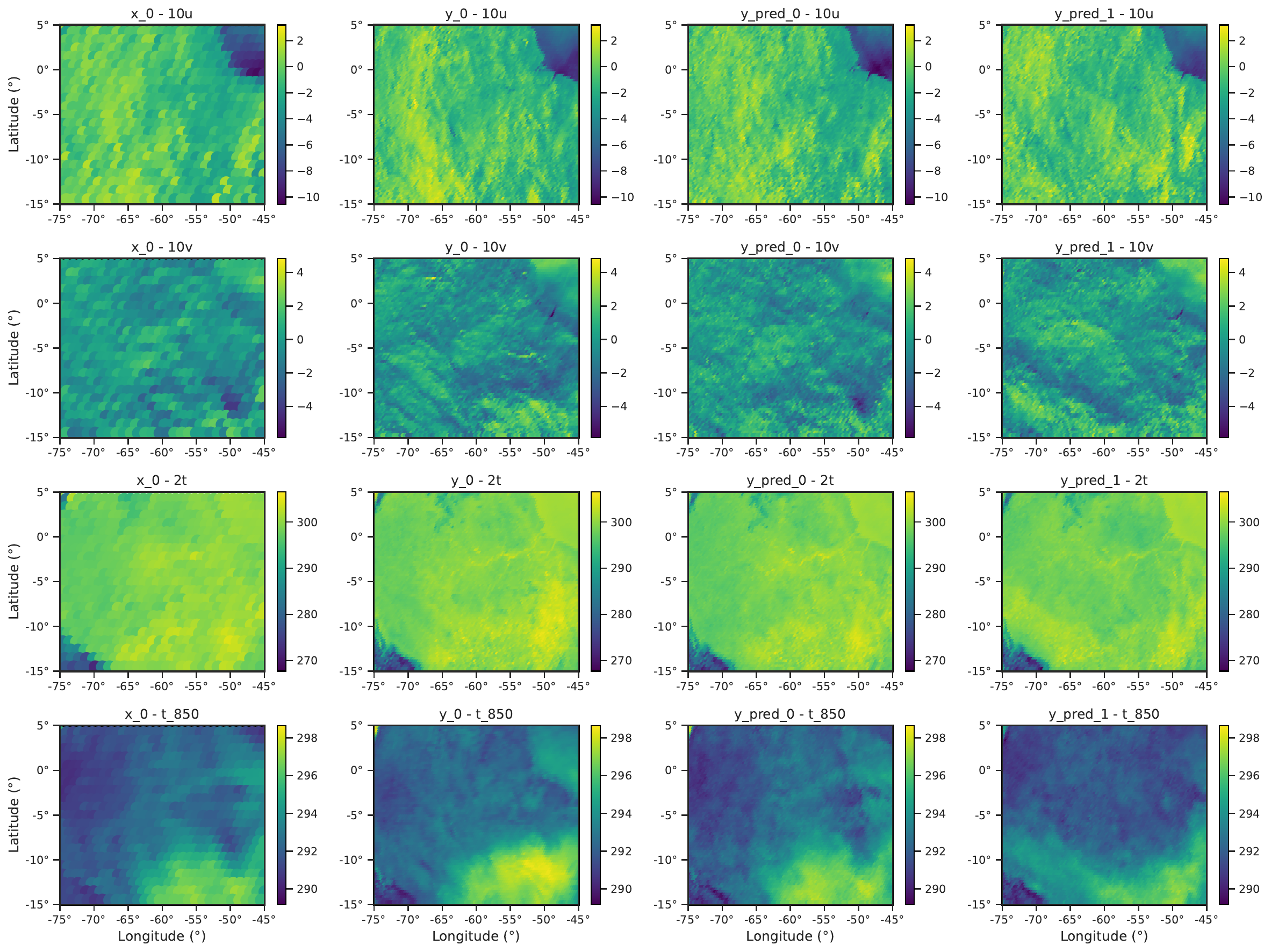}
    \caption{Subseasonal ensemble forecast first member (first column), medium-range ensemble forecast first member (second column), and downscaling with the diffusion model of two subseasonal members (third and fourth columns, with the third corresponding to downscaling of the first column). Results are shown for the Amazon rainforest at lead time 24h. Rows display 10u, 10v, 2t, and T850. The diffusion model generates stochastic and physically consistent small-scale features. }
    \label{res:spectra:amazon:fig}
\end{figure}
Figure~\ref{res:spectra:amazon:fig} illustrates the role of probabilistic downscaling in a weakly  constrained region, the Amazon rainforest, where small-scale structures are only weakly determined by the coarse-scale state. In such areas, a deterministic model trained with an RMSE objective tends to produce overly smooth fields and fails to recover realistic fine-scale variability. By contrast, the diffusion model generates diverse small-scale structures while remaining physically consistent with the large-scale flow.

\begin{figure}
    \centering
    \includegraphics[width=1\linewidth]{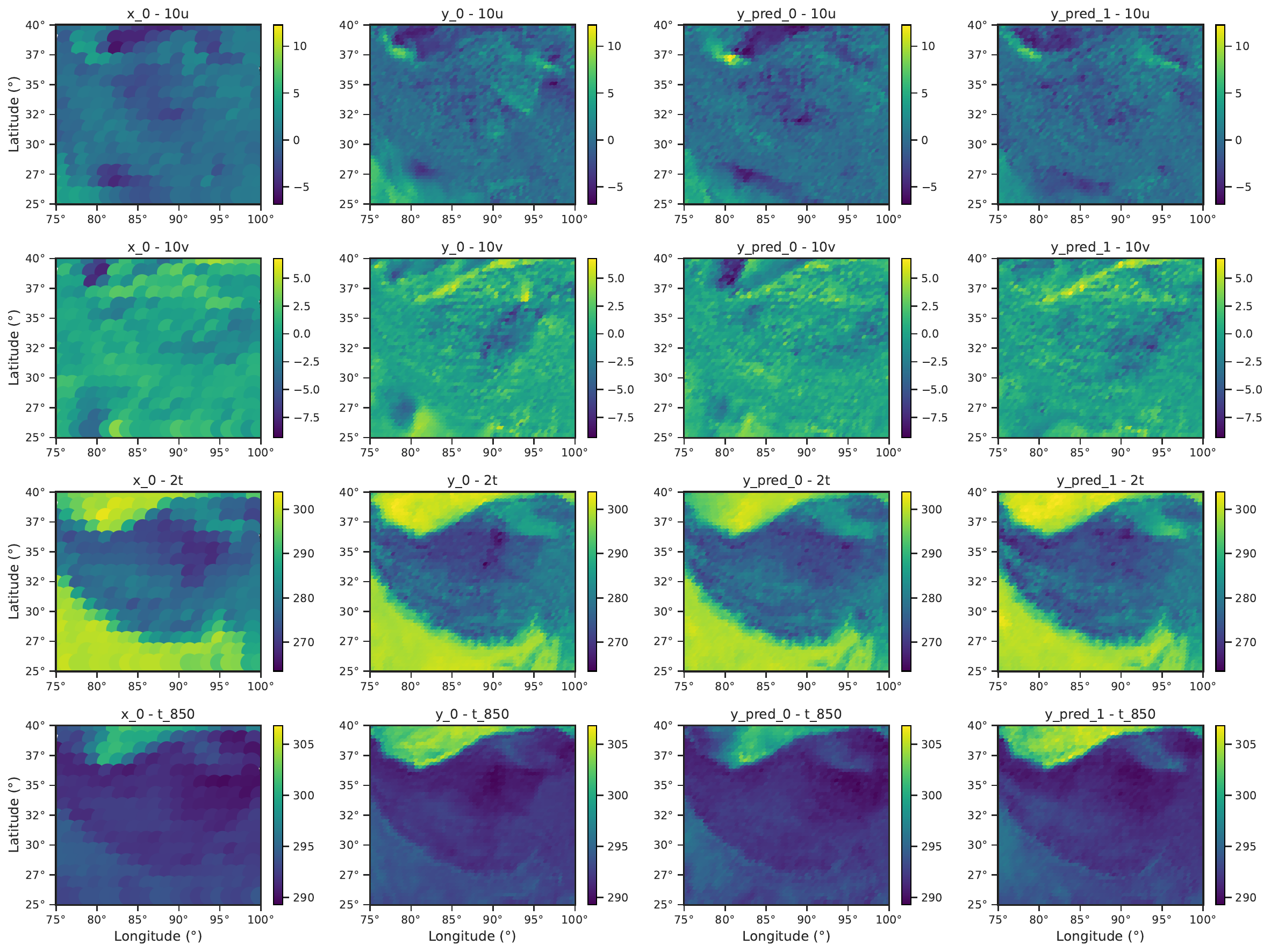}
    \caption{Subseasonal ensemble forecast first member (first column), medium-range ensemble forecast first member (second column), and downscaling with the diffusion model of two subseasonal members (third and fourth columns, with the third corresponding to downscaling of the first column). Results are shown for the Himalayas at lead time 24h. Rows display 10u, 10v, 2t, and T850. The diffusion model generates physically consistent small-scale features.}
    \label{res:spectra:himalayas:fig}
\end{figure}

Figure~\ref{res:spectra:himalayas:fig} shows a contrasting regime over the Himalayas, where small-scale features are much more strongly constrained by the large-scale state and by topography. In this case, the difference between deterministic and probabilistic downscaling is limited (not shown), as much of the fine-scale structure is already determined by the coarse-resolution input. Similar behaviour is obtained even when orography is removed from the input, suggesting that its influence can largely be inferred indirectly from other variables, such as near-surface temperature.

\begin{figure}
    \centering
    \includegraphics[width=1\linewidth]{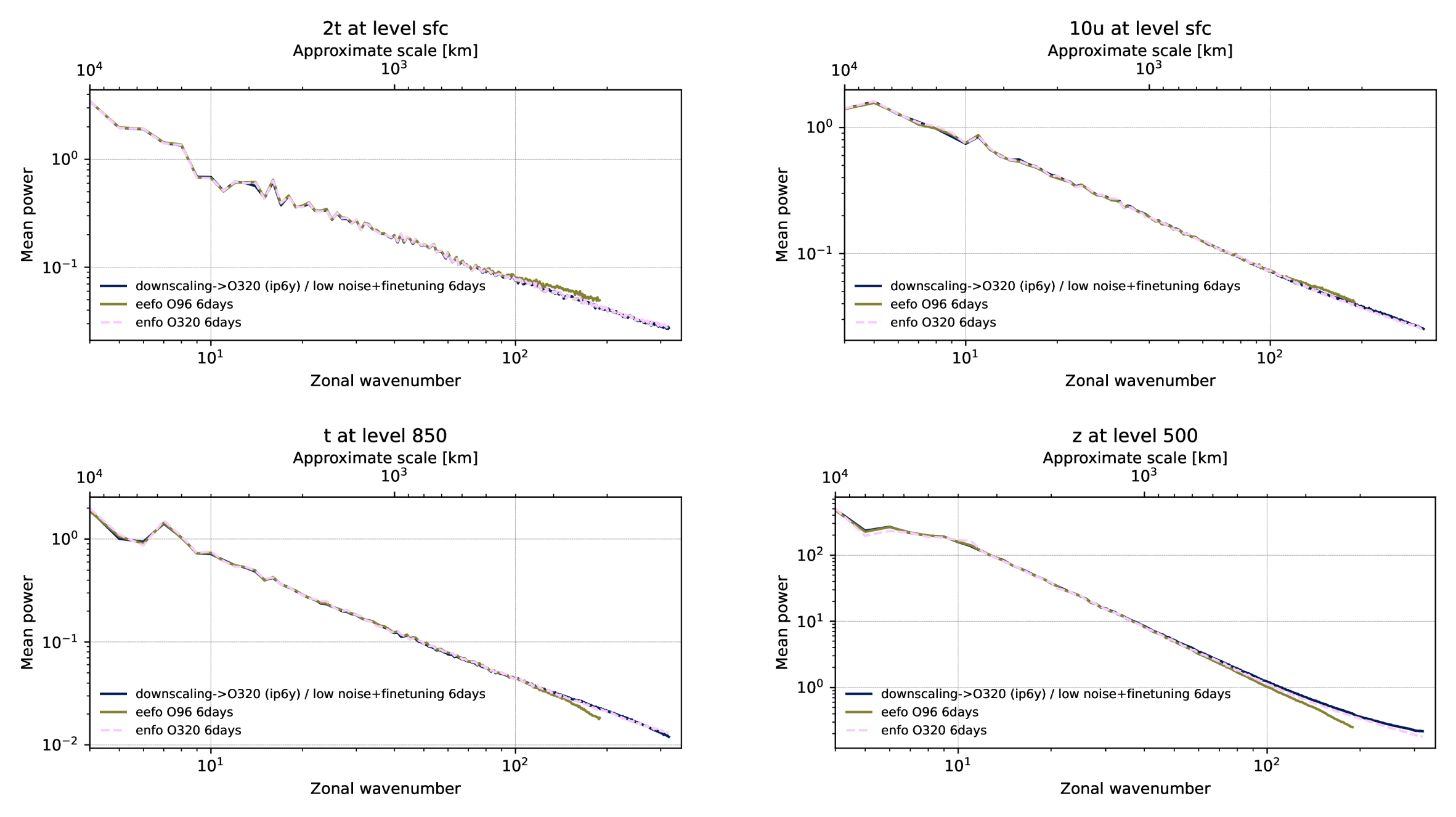}
    \caption{Comparison of global amplitude spectra of weather variables from the subseasonal forecasts, the medium-range forecasts, and the downscaled subseasonal forecasts. The spectra are evaluated over the full globe at forecast day 6 and are based on 31 forecast initialisation dates and 10 ensemble members per date, corresponding to a sample of 310 forecasts.}
    \label{res:spectra:spectra:fig}
\end{figure}

Figure \ref{res:spectra:spectra:fig} compares spectra of the input, the downscaled input, and the target. The diffusion model output departs from the subseasonal input spectrum and aligns with the medium-range forecast target. Except for z500, the small-scale zonal wavenumber powers of the downscaled ensemble match those of the target. The slight mismatch at small scales for z500 may result from the loss function emphasis on surface variables.

\subsection{Case studies: tropical cyclones}
\label{res:TC}

Downscaling is particularly relevant for recovering extremes, such as tropical cyclones (TCs), where high-resolution fields contain extreme values absent at lower resolutions. For TCs, this concerns minimum mean sea level pressure (in the eye of the cyclone) and maximum wind speed around the eye of the cyclone. We examine two TCs where the pdfs of these extremes differ markedly between subseasonal and medium-range forecasts, to assess whether downscaling can generate extreme values consistent with the medium-range forecasts.

\subsubsection{Idalia}
\label{res:TC:idalia}

We first consider Tropical Cyclone Idalia, which affected the eastern United States in late August 2023 and reached an observed minimum central pressure of approximately 942\,hPa.
Because our analysis is performed on coarsened ensemble forecasts, neither the downscaled fields nor the target fields are expected to reproduce the observed minimum pressure itself.
The aim is instead to assess whether the distributions of mean sea-level pressure (MSLP) and near-surface wind speed produced by the downscaling are consistent with those of the target ensemble.

\begin{figure}
    \centering
    \includegraphics[width=1\linewidth]{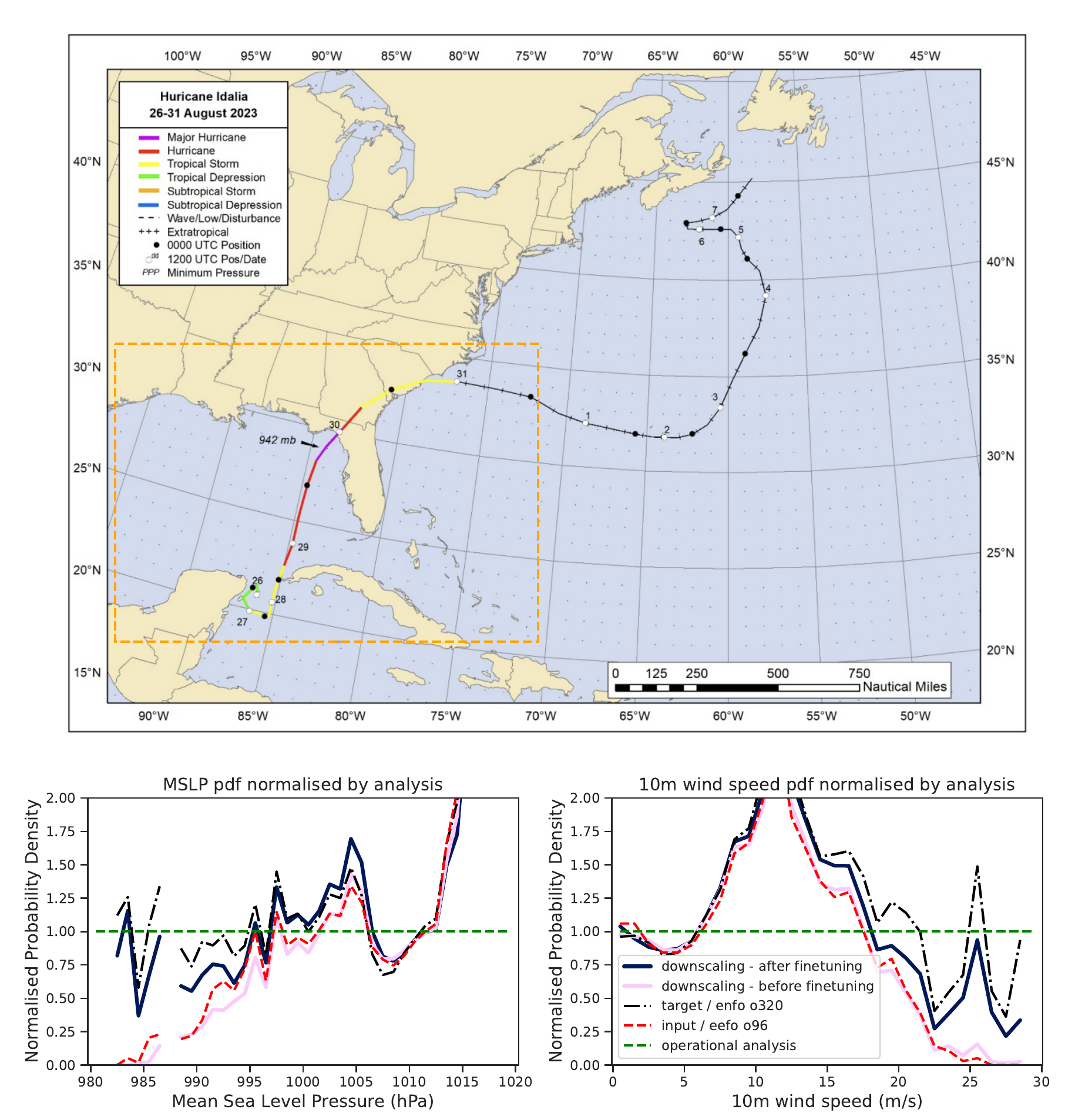}
    \caption{Normalised pdfs of mean sea level pressure (980–1015 hPa) and 10-meter wind speed (0–30 m/s) for Hurricane Idalia across ensemble members, initialized on 26-30 August 2023, with a 1 to 5-days lead time. The pdfs represent the distributions of gridpoint values within the cyclone region, normalised by the corresponding distribution of the operational analysis.
    They are shown for the target IFS medium-range ensemble forecasts, the input IFS subseasonal ensemble forecasts, and the downscaled reconstructions (before and after finetuning).
    The accompanying map illustrates Idalia’s track from 27 to 31 August 2023, marking stages from tropical depression to major hurricane.
    Idalia cyclone track is adapted from NOAA/NHC \url{https://www.nhc.noaa.gov/data/tcr/index.php?season=2023&basin=atl}.
    }
    \label{res:TC:idalia:pdf:fig}
\end{figure}

Figure~\ref{res:TC:idalia:pdf:fig} compares the distributions of gridpoint mean sea level pressure and 10-m wind speed within the cyclone region, expressed as normalised probability density functions, for two downscaling configurations: (i) no finetuning, sampled with a maximum noise of 80; (ii) finetuned on the high-noise distribution, sampled with a maximum noise of 100{,}000, alongside the input and the target (Section~\ref{model:training}).

Without finetuning, the model fails to capture the lowest MSLP and the highest wind speeds.
With finetuning and very high sampling noise, the downscaled PDFs match the target tails and clearly diverge from the input at extremes, with a slight underestimation of peak wind speed.

\begin{figure}
    \centering
    \includegraphics[width=1\linewidth]{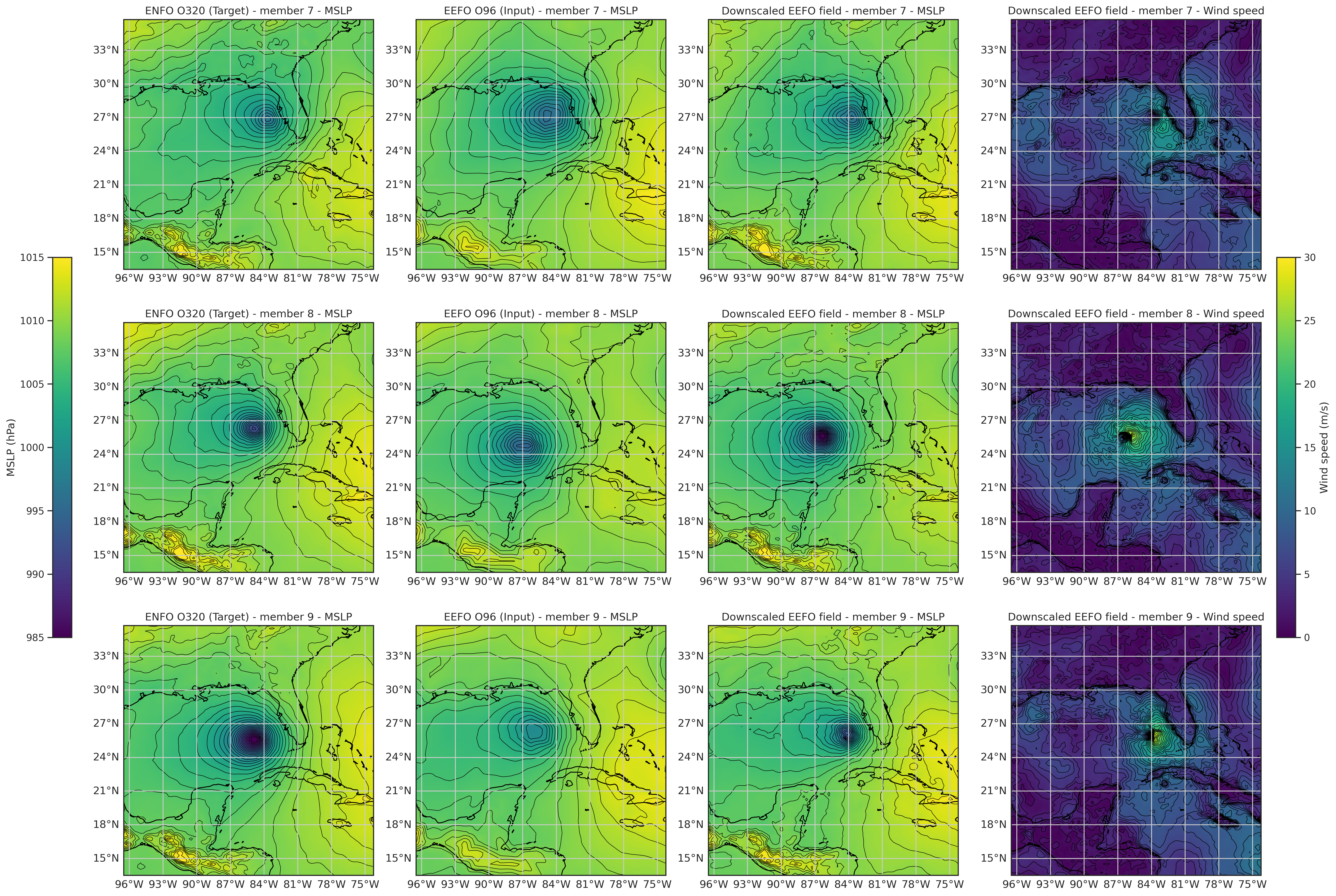}
    \caption{Mean sea level pressure (985–1015 hPa) and wind speed (0–30 m/s) for tropical cyclone Idalia for three ensemble members at 24-hour lead time, initialized on 28 August 2023. The first column shows the target IFS medium-range ensemble forecast MSLP, the second column displays the input IFS subseasonal ensemble forecast MSLP, and the third and fourth columns present the MSLP and wind fields, respectively, reconstructed through downscaling of the IFS subseasonal ensemble forecast. }
    \label{res:TC:idalia:fields:fig}
\end{figure}

Figure~\ref{res:TC:idalia:fields:fig} shows that downscaling can both increase the intensity of the tropical cyclone and relocate it. Despite this displacement, the fields remain dynamically consistent: the MSLP minimum co-locates with the circulation centre, and the wind structure organises consistent with the isobars.

\subsubsection{Hilary}
\label{res:TC:hilary}

The second case is Hurricane Hilary, which affected the west coast of Mexico in mid-August 2023 and reached a minimum central pressure of about 940\,hPa.
As in the Idalia case, the objective is not to reproduce the observed minimum exactly, but to assess whether the downscaled ensemble recovers the intensity range and distributional shape of the target ensemble.

\begin{figure}
    \centering
    \includegraphics[width=1\linewidth]{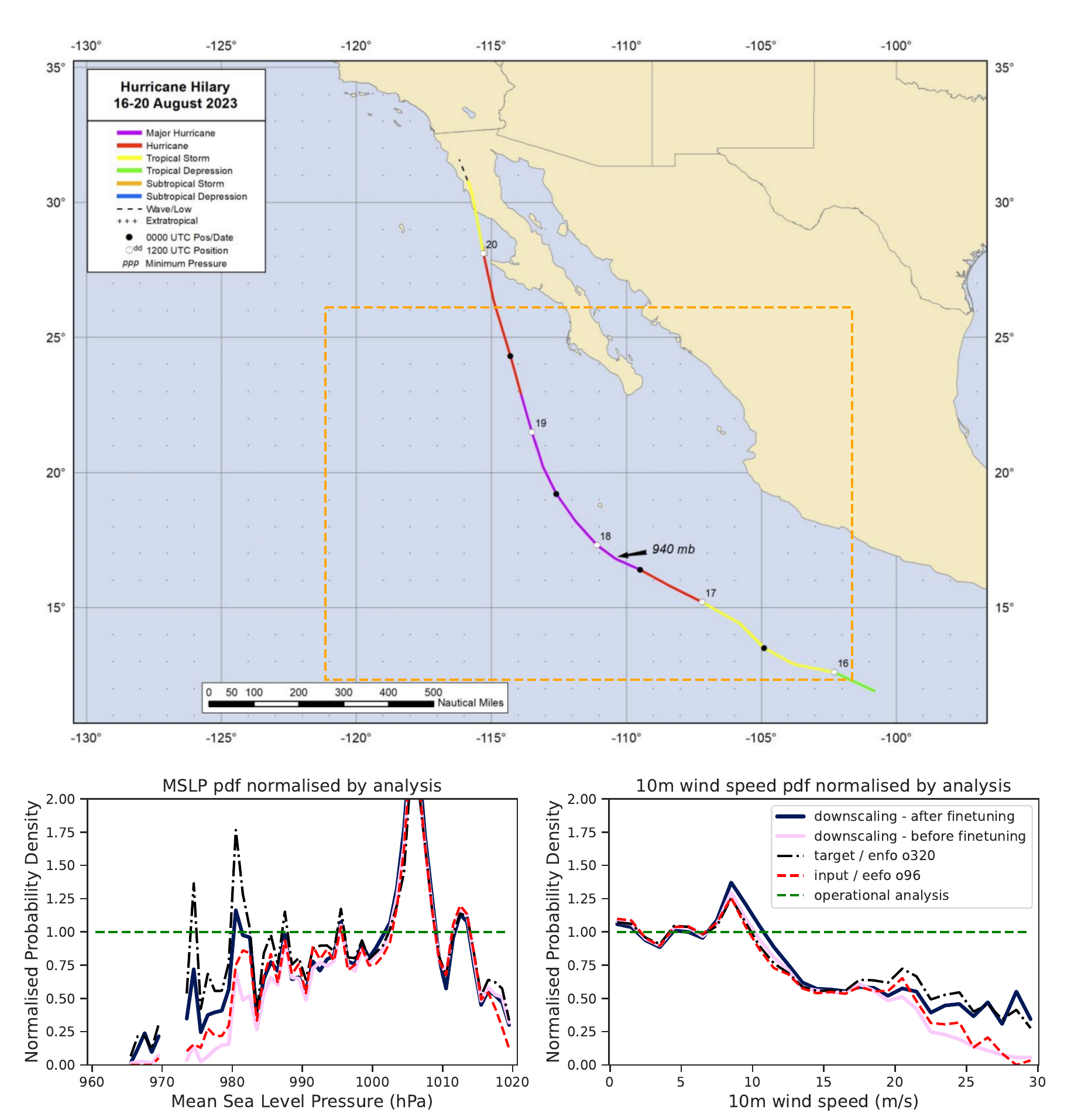}
    \caption{Normalised pdfs of mean sea level pressure (960–1015 hPa) and 10-meter wind speed (0–30 m/s) for Hurricane Hilary across ensemble members, initialized on 16-20 August 2023, with a 1 to 5-days lead time. The pdfs represent the distributions of gridpoint values within the cyclone region (approximately the dashed orange line), normalised by the corresponding distribution of the operational analysis.
    They are shown for the target IFS medium-range ensemble forecasts, the input IFS subseasonal ensemble forecasts, and the downscaled reconstructions (before and after finetuning).
    The map illustrates the cyclone’s track from 16 to 20 August 2023, marking stages from tropical depression to major hurricane.
    Hilary cyclone track is adapted from NOAA/NHC \url{https://www.nhc.noaa.gov/data/tcr/index.php?season=2023&basin=epac}.
    }
    \label{res:TC:hilary:pdf:fig}
\end{figure}

\begin{figure}
    \centering
    \includegraphics[width=1\linewidth]{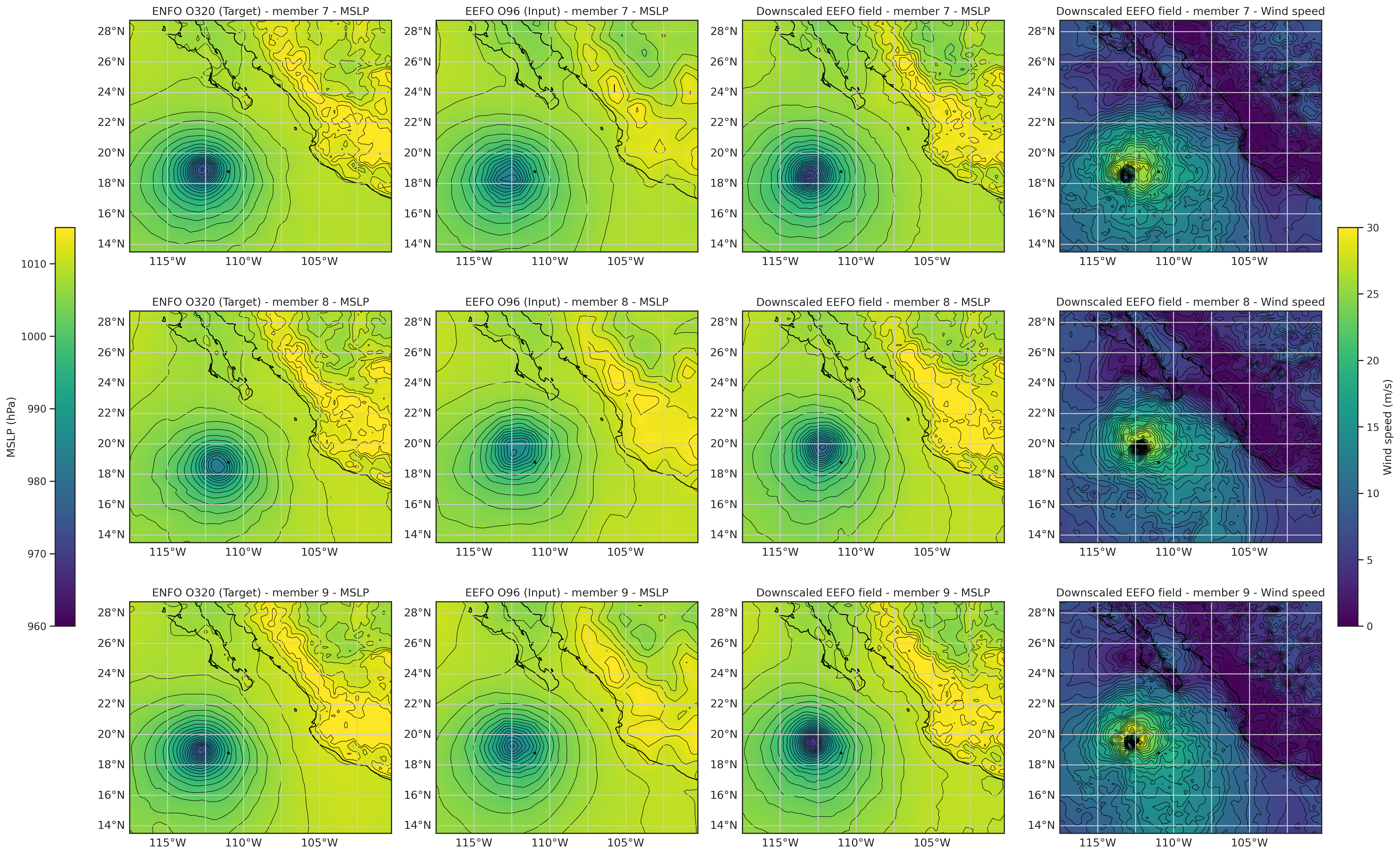}
    \caption{Local fields of mean sea level pressure (960–1015 hPa) and wind speed (0–30 m/s) for Tropical Cyclone Hilary across ensemble members at 24-hour lead time, initialized on 17 August 2023. The first column shows the target IFS medium-range ensemble forecast MSLP, the second column displays the input IFS subseasonal ensemble forecast MSLP, and the third and fourth columns present the MSLP and wind fields, respectively, reconstructed through downscaling of the IFS subseasonal ensemble forecast.}
    \label{res:TC:hilary:fields:fig}
\end{figure}

As shown in Figure~\ref{res:TC:hilary:pdf:fig}, the downscaled ensemble reaches MSLP values of approximately 960\,hPa and, after high-noise fine-tuning, its pressure and wind distributions become closely aligned with those of the target ensemble, including in the lower-pressure tail.
Figure~\ref{res:TC:hilary:fields:fig} shows that this improvement is also reflected in the spatial fields, with a more intense and coherent cyclone structure than in the input forecasts.

This behaviour is consistent with results obtained for five tropical cyclones from August 2023---Franklin, Idalia, Hilary, Fernanda, and Dora---for which high-noise fine-tuning systematically improves the representation of the low-pressure and high-wind tails.

\section{Conclusions and perspectives}
\label{conclu}

We introduce \aifsdd{}, a probabilistic diffusion-based framework for ensemble downscaling. It transforms low-resolution ensemble forecasts into high-resolution ensembles by learning conditional distributions of fine-scale residuals. The approach builds on AIFS-Diffusion, combining the AIFS graph encoder–processor–decoder with a diffusion model inspired by EDM principles \citep{karras_elucidating_2022}. 
This work is intended as a step towards operational kilometre-scale ensemble forecasting, by exploring diffusion-based downscaling as an efficient and scalable route to high-resolution ensemble generation, notably for Destination Earth, the EU initiative to build digital twins of the Earth system.

We evaluate the method on the 100\,km~$\rightarrow$~30\,km downscaling task.
The evaluation combines physical-realism diagnostics, spectral analyses, probabilistic skill scores, and tropical-cyclone case studies.
The downscaled ensemble reproduces small-scale spectral energy, preserves key multivariate relationships such as wind–pressure coupling in cyclones, and shows encouraging surface-variable FCRPS, but it remains somewhat overdispersive relative to the medium-range ensemble.
Because EDM was originally designed for images, its application to weather fields requires tailored adaptations to adequately explore the full high-resolution distribution. In particular, capturing long-tail/extremes behaviour requires finetuning in high-noise regimes: models trained only at low noise under the EDM lognormal schedule produce realistic structures and good probabilistic skill but underrepresent extremes.
Inference is computationally efficient: a 15-day member at 24-hour output frequency takes roughly four minutes on a single A100 40 GB GPU, including I/O, enabling large global ensembles. The method applies to both regional and global configurations.

Future developments will focus on extending the method and broadening its applicability. Scaling to higher spatial resolutions is a natural next step, enabling the generation  of probabilistic forecasts at higher resolutions than   those computationally affordable operationally with physical models and large ensemble sizes ($\ge 50$ members). 
Temporal extensions are planned to move from independent snapshots towards 4D generation, ensuring physically consistent evolution of the small-scale features across lead times. The integration within the Anemoi framework will enable large-scale, real-time production of high-resolution ensemble forecasts.
Beyond weather ensemble forecasts, the same framework could be applied to climate downscaling, offering a cost-effective alternative to dynamical approaches.

\subsection*{Acknowledgements}

Destination Earth is implemented under the leadership of the European Commission’s DG CNECT, in partnership with the European Space Agency and EUMETSAT. We acknowledge the support and funding provided through this programme. We also thank the EuroHPC Joint Undertaking for granting access to the EuroHPC systems Leonardo and MareNostrum 5. We are grateful to Tobias Finn, Mihai Alexe, the ECMWF AIFS team, the many colleagues at the ECMWF who supported this work, and Météo-France and MET Norway for the meaningful discussions. We further acknowledge the support of Peter Düben and Irina Sandu throughout the project.

\bibliographystyle{plainnat} 
\bibliography{paper_downscaling.bib}

\end{document}